\documentclass[twocolumn,showpacs,preprintnumbers,amsmath,amssymb]{revtex4}
\usepackage{amsmath,amsthm}
\usepackage{graphicx}
\usepackage{dcolumn}
\usepackage{bm}

\begin{document}
\title{Isospin Effects on Astrophysical S-Factors} 
\author{Sachie Kimura}
\author{Aldo Bonasera}
 \altaffiliation[Also at ]{Libera Universit\`{a} Kore di Enna, 94100 Enna, Italy}
 \affiliation{Laboratori Nazionali del Sud, INFN,
via Santa Sofia, 62, 95123 Catania, Italy}

\date{\today}

\begin{abstract}
We estimate the ratios of bare astrophysical S-factors at zero incident energy for 
proton and deuteron induced reactions in a model 
which assumes a compound nucleus formation probability plus a statistical decay.  
The obtained ratios agree well with available experimental values, as far as 
the reactions which have dominant $s$-wave entrance channel components are investigated.
Due to its simplicity the model could be used as a guidance for predictions on reactions
which have not been investigated yet.
\end{abstract}
\maketitle

The nuclear fusion cross sections of proton and deuteron induced reactions at low energies are 
of particular interest from the points of view of 
the stellar nucleosynthesis and the nuclear energy production.     
These cross sections are measured at laboratory energies and extrapolated 
to thermal energies~\cite{nacre,yan_prc}, because of their small values at such low energies. 
This extrapolation is done by introducing the astrophysical S-factor:
\begin{equation}
  \label{eq:sf}
  S(E)=\sigma(E)E e^{2\pi \eta(E)},
\end{equation}
where $\sigma(E)$ is the reaction cross section at the incident center-of-mass energy $E$ 
and $\eta(E)=Z_T Z_P \alpha \sqrt{\frac{\mu c^2}{2E}}$, $Z_T$, $Z_P$, $\mu$ denoting the 
atomic numbers and the reduced mass of the target and the projectile. $\alpha$ and $c$ are 
the fine-structure constant and the speed of light, respectively.   
The exponential term in the equation represents the Coulomb barrier penetrability.  
Since one has factored out the strong energy dependence of $\sigma(E)$ due to the barrier 
penetrability, the S-factor could be approximated by a smooth energy dependence in the absence 
of low-energy resonance.  
A lot of effort has been made to extract the S-factor in the low energy region 
in particular for transfer reactions experimentally~\cite{erag,thm2,la01,bemmerer:122502}.
Although to know the low energy S-factor for radiative capture reactions, still one needs the 
extrapolation with theoretical formulae~\cite{akram}.  
Owing to this experimental effort
one can, in principle, determine the S-factor in the energy region of astrophysical interest directly. 
However at such a low energy it is known that electrons around the target nucleus have an 
effect on the fusion cross section.  
In contrast, in the stellar nucleosynthesis, nuclei are almost fully ionized and are surrounded by 
the plasma electrons. The nuclear reactions in such a circumstance are affected by a different 
mechanism of the plasma electron screening~\cite{ichimaru,shav}.         
Hence the screening effects of the bound electrons should be removed from the S-factor data, 
in order to asses the reaction rate in the stellar site correctly.
The enhancement by the bound electrons 
is discussed in terms of a constant potential shift(screening potential $U_e$)
~\cite{skls,frvr,la01,thm2,kb-cdf,kb-icfe}.

In this paper we would like to investigate in detail the physical origins of the bare S-factor.
We propose visualizing the fusion mechanism under the assumption of the compound nucleus (CN) 
formation as a two step process:

STEP 1) The nuclear fusion under the Coulomb barrier occurs with a given probability which depends crucially on nuclear effects 
(the nuclear potential, nuclear sizes, an excitation of collective modes etc..).  A CN is formed at a given excitation energy.  
Since we are interested in determining the S-factor in the limit of zero bombarding energies, the excitation energy is essentially 
given by the Q-value of the reaction which leads to the CN.

STEP 2) The CN decays into a given channel.  We assume this decay process to be statistical and make use of 
the Weisskopf model \cite{weiss,bon87} to evaluate its probability.

\begin{figure}
  \begin{center}
  \includegraphics[height=.25\textheight]{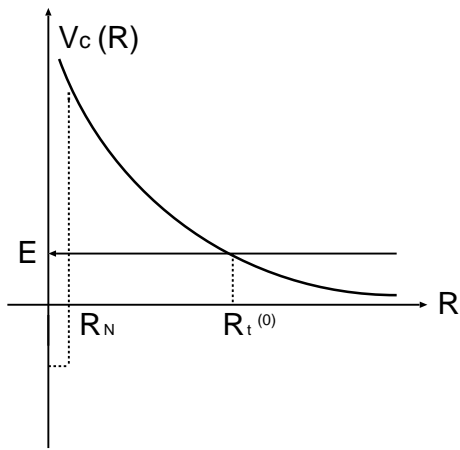}    
  \end{center}
  \caption{A schematic feature of the interaction between colliding nuclei.}
  \label{fig:fig1}
\end{figure}

Low energy behaviors of the S-factors have been studied by means of 
the R-matrix theory~\cite{ad,daacv,barker}, which is also based on a description of the nuclear reaction through 
the intermediate CN.  
In the nuclear reactions in general, it is known that 
this CN process coexists with the direct process mechanism~\cite{thomas}.  
Provided that the energy regions of our interest in this paper is low, 
we presume that the $l=0$ partial wave component would be dominant for most of the reactions
investigated. This fact suggests that most of low energy reactions are possibly to be described by using only 
the CN mechanism. Certainly there are some reactions which do not have $l=$0 component, these cases will be discussed separately.
The situation of our interest is displayed in Fig.~\ref{fig:fig1}, 
where the colliding ions at a given beam energy $E$ in the center of mass (CM) system must 
overcome the Coulomb barrier in order to fuse.
In general, the cross section for the sub-barrier fusion at energy $E$ and angular momentum $l$ is given by~\cite{bk1}:
\begin{equation}
  \label{eq:sfs}
  \sigma(E)=\pi \hbar^2 /(2\mu E) \sum_{l=0} (2l+1)T_l(E),
\end{equation}
where $T_l(E)$ is the tunneling probability.  
We take the limit of $E\rightarrow 0$ 
and keep only $l=0$ term of the sum in Eq.~(\ref{eq:sfs}) in the following discussion. 
However one can easily extend the case where the centrifugal potential term is included~\cite{clayton} and later we will only mention the modification by the $l \ge 1$ terms to the $l=0$ case.  
We point out that $T_l(E)$ is not the pure Coulomb penetrability but the penetrability of the Coulomb + 
centrifugal + nuclear potentials.
The tunneling probability $T_0(E)$ can be written in terms of the action ${\cal A}$ in the low incident energy limit~\cite{clayton,bk1}:
\begin{equation}
  T_0(E) \sim \exp\left[-\frac{2}{\hbar} {\cal A} \right], \hspace*{0.5cm} {\cal A} = \int^{R_t^{(0)}}_{R_N} dr \sqrt{2\mu [V_c(r)-E]}, \label{eq:peneta} 
\end{equation}
where $V_c(r), R_N$ and $R_t^{(0)} $ are the Coulomb potential, the inner and the outer classical turning points,
respectively, as they are shown in Fig.~\ref{fig:fig1}. 
In particular, if we consider the pure Coulomb penetrability, i.e. we take the limit of $R_N \rightarrow 0$,   
${\cal A}$ becomes ${\cal A}_G=Z_TZ_Pe^2\sqrt{\frac{2\mu}{E}}$: in the Gamow limit. As a consequence Eq.~(\ref{eq:peneta}) reduces to the Gamow factor which is exactly the inverse of the exponential term of 
Eq.~(\ref{eq:sf}).
Comparing eqs.~(\ref{eq:sf}) and (\ref{eq:sfs}) we derive an S-factor in the Gamow limit: $S_G= \pi \hbar^2 /(2\mu).$
It means that the S factor would be independent of the energy and of the exit channels, 
if the process would be dominated by Coulomb.
However the actual value of the $S$-factor depends on the colliding ions and on the exit channels.
For instance let us consider two reactions $^{10}$B($d,\alpha)^8$Be ($S_{exp}=102.7$ MeV b)~\cite{yan} 
and $^{11}$B($p,\alpha)^8$Be ($S_{exp}=208.0$ MeV b)~\cite{aerrss} which have identical exit channels
but different entrance channels.  Our simple formula gives $S_G=0.39$ and 0.71 MeVb, respectively, which 
disagree with both experiments. 
This simple comparison suggests that the nuclear effects are really dominant and should be taken into account.
A modification due to the nuclear effects can be easily included assuming that the nuclear potential is given by a square 
well of radius $R_N$ indicated by the dotted line in Fig.~\ref{fig:fig1}~\cite{clayton}.  Here we will assume the empirical
formula: $R_N=\alpha_n R_{No}$, where $R_{No}=1.2 \times (A_P^{1/3}+A_T^{1/3})$ [fm] is the sum of 
the radii of the projectile and the target, denoting the mass numbers of the target and the projectile nuclei 
by $A_T$ and $A_P$, respectively.  $\alpha_n$ is a parameter 
which takes into account the fact that nuclei have a diffuse potential and during the reaction some collective
degrees of freedom might be excited which could modify the effective radii of the ions.  

In the approximation considered, the action is written down in terms of ${\cal A}_G$ as:
\begin{equation}
  \label{eq:sfn}
  {\cal A}= {\cal A}_G-\int_0^{R_N} dr \sqrt{2 \mu\left[V_c(r)
      -E\right]},
\end{equation}
where the integral of the right hand side is the action of the pure Coulomb potential from the origin to the nuclear interaction radius.   
This action integral is approximated by $2\sqrt{2\mu Z_TZ_Pe^2R_N}$, 
in the limit of $E\rightarrow 0$ and therefore the S-factor modified by nuclear effects is written by, 
  \begin{equation}
  \label{eq:sfN}
  S_{No}= S_G e^{\frac{4}{\hbar} \sqrt {2\mu Z_T Z_P R_N}}.
\end{equation}
A simple inspection of Eq.~(\ref{eq:sfN}) using standard values is not able to reproduce the experimental S-factors 
for the two reactions discussed above.  
This implies that 
the modification is not enough, in fact we have only included STEP 1) of our proposed scenario.
We need now to multiply $S_{No}$ by the probability of getting a certain exit channel from the CN $\Pi(A^*\rightarrow c+d)$.
The Weisskopf model\cite{weiss} gives an expression for the probability to obtain a certain exit channel $i$ as a function of 
the cross section of the reverse process $\sigma^{abs}_i$:
\begin{equation}
  \label{eq:sfW}
  \Pi_i= g_i(T_{k_i}^2+2m_iT_{k_i})\frac{\rho_i(E_{CN}^*-T_{k_i})}{\rho(E_{CN}^*)}\sigma^{abs}_i.
\end{equation}
One can, therefore, express the probability $\Pi$ as $\Pi _i /\sum \Pi _i $,
where the sum in the denominator is taken over all possible exit channels. 
In Eq.~(\ref{eq:sfW})  
$g_i$ denotes the number of states for the spin of the particle considered. 
($g_i=2s_i+1$, where $s_i$ is the spin for proton, neutron and alpha particle, $g_i=2$ for $\gamma$~\cite{bon87}.)  
$T_{k_i}$ and $m_i$ are the kinetic energy and the mass, respectively, of the lightest reaction product 
in the exit channel. 
$\rho_i(E_{CN}^*-T_{k_i})$ and $\rho(E_{CN}^*)$ are the level densities of the heavier reaction product and of the CN, 
respectively, where $E_{CN}^*$ is the excitation energy of the CN. Introducing the entropy of the nuclei, 
the ratio of $\rho_i(E_{CN}^*-T_{k_i})$ to $\rho(E_{CN}^*)$ can be written as
\begin{equation}
  \label{eq:rtw}
  \frac{\rho_i(E_{CN}^*-T_{k_i})}{\rho(E_{CN}^*)}=\exp(-\frac{T_{k_i}}{T}), 
\end{equation}
where $T$ is the temperature of the CN, in the unit of energy. 
The CN temperature is given by $T=\sqrt{Q_{CN}/a}$, $Q_{CN}$ is the Q-value for the
CN formation and $a\sim A_{CN}/8.0$ [MeV$^{-1}$] is the level density parameter~\cite{weiss}.  
Here we have denoted the mass number of the CN to be $A_{CN}$ and have assumed that 
all the excitation energy is given by the Q-values (i.e.$E \rightarrow 0$).
This is the required STEP 2) of our model. We can, therefore, define
a S-factor of the $l=0$ entrance channel as:
\begin{equation}
 \label{eq:sfN2}
S_N= S_G e^{\frac{4}{\hbar} \sqrt{2\mu Z_TZ_P R_N}} \Pi.
\end{equation}
The factor depends on the exit channel as expected. 
In passing we mention that the $S$-factor of the reactions which have $l\ge 1$ partial wave components
in addition to the $l=0$ component can be written by~\cite{clayton}
\begin{equation}
 \label{eq:sfN3}
S_N= S_G e^{\frac{4}{\hbar}\sqrt{2\mu Z_TZ_P R_N}}\left[1+\sum_{l\ge 1}(2l+1)e^{-\frac{2l(l+1)\hbar}{\sqrt{2\mu Z_TZ_P R_N}}}\right] \Pi.
\end{equation}      
One can, in principle, obtain the absolute value of the S-factor from Eq.~(\ref{eq:sfN2}), however 
to determine the level density $\rho(E_{CN}^*)$ one needs some numerical calculations.  
We will, therefore, discuss 
limiting cases where the model gives an analytical solution, in order to get a feeling of the proposed approach.  
The first case is given by the ratios of S-factors (in the limit of zero beam energy) where the CN formed is the same 
but two different reaction channels are studied (different isospins in the exit channel), while the second is the opposite i.e. different isospin in the
entrance channel and the same decay mode.  In this way, by taking ratios, screening effects should exactly cancel in both cases.

When the two reactions with an identical entrance channel are considered, its S-factor ratio $R$ is given by:
  \begin{equation}
 \label{eq:sfR}
R=\frac{g_j(T_{k_j}^2+2m_jT_{k_j})\sigma_j^{abs}}{g_i(T_{k_i}^2+2m_iT_{k_i})\sigma_i^{abs}} e^{-\Delta Q/T}
\end{equation}
where $j (i)$ refers to the decay channels, and $T_{k_{j(i)}}=\frac{A_2}{A_1+A_2}Q_{j(i)}$.     
The reaction Q-value for each channel and its difference is denoted by $Q_{j (i)}$ and $\Delta Q=Q_j-Q_i$, respectively.  
We write the mass numbers of particles in the exit channels as $A_1$ for the lightest particle and $A_2$ for the rest.    
Thus in the defined ratio the 'entrance channel' effects drop out. In the case of the radiative capture the equation above can be easily 
modified using the relativistic Weisskopf 
equation given above, eq.(\ref{eq:sfW}).  The absorption cross sections are taken to be geometrical but modified by a Coulomb barrier:
\begin{equation}
  \label{eq:cc}
  \sigma^{abs}_j=\pi R_N^2(1-V_c(R_N)/Q_j),
\end{equation}
where $R_N$ is the sum of the radii of the nuclei in the exit channel $j$.
For the photo-absorption cross section we make use of the parameterization given in \cite{Ahr}. 
\begin{equation}
  \label{eq:pac}
  \sigma^{abs}_j=75A_{CN}\frac{(T_{k\gamma}-2.226)^{3/2}}{T_{k\gamma}^3} [mb],
\end{equation}
where $T_{k\gamma}=Q_{CN}$. This parameterization holds well for relatively high $\gamma$ energies ($T_{k\gamma} > 23$MeV).
We will make use of it, if there is no experimental data available at lower energies.

Within this simple approach we have estimated the S-factor ratios for the proton and deuteron induced reactions
with positive Q-values on the nuclei D, $^6$Li, $^7$Li, $^9$Be, $^{10}$B and $^{11}$B.
All the investigated reactions are listed in Table~\ref{tab:a} and~\ref{tab:b}. 
We take the ratios of the 2nd to the 3rd column reactions. 
Their S-factor ratios determined from experimental data are shown in the 4th column
and are compared with 
the ratios from Eq.~(\ref{eq:sfR}) in the last column.    
In Table~\ref{tab:b} we made a list of the ratios of the radiative capture reactions to the 
transfer reactions. 
From the tables we see a general good agreement of our estimations to data.  
\begin{table}
\caption{Ratios of $S$-factors for the transfer reactions with identical entrance channels and different exit channels. }
\label{tab:a}
\begin{ruledtabular}
\begin{tabular}{cccc}
 \multicolumn{2}{c}{$A_{CN}$}      & &   \\
 Reaction1 & Reaction2    &  Ratio & Ratio   \\
 $S_0$(MeVb) & $S_0$(MeVb) & (data) & (calc.) \\
\hline
 \multicolumn{2}{c}{$4$}     &   &   \\
 $d$($d,p$)$t$ & $d$($d,n$)$^3$He  &   &    \\
0.0571~\cite{daacv} & 0.0524~\cite{daacv} & 1.09 & 0.97 \\ 
 & & & \\
\multicolumn{2}{c}{$8$}  &   &    \\
$^6$Li($d,n$)$^7$Be & $^6$Li($d,p$)$^7$Li  &   &    \\
25.3 & 26.6~\cite{czerski} & 0.95~\cite{czerski} & 1.26 \\ 
 & & & \\
$^6$Li($d,\alpha$)$^4$He & $^6$Li($d,p$)$^7$Li   &   &    \\
23.1~\cite{PhysRevC.55.1517} & 26.6~\cite{czerski} & 0.87 & 0.37   \\ 
 & & & \\
\multicolumn{2}{c}{$10$}  &   &    \\
$^9$Be($p,d$)$^8$Be & $^9$Be($p,\alpha$)$^6$Li  &   &    \\
18.~\cite{nacre} & 18.~\cite{nacre} & 1.0 & 0.95  \\ 
 & & & \\
\multicolumn{2}{c}{$11$}  &   &    \\
$^9$Be($d,p$)$^{10}$Be & $^9$Be($d,\alpha$)$^7$Li  &   &    \\
15.2~\cite{yan,yan_prc} & 20.8~\cite{yan,yan_prc} & 0.73 & 0.79  \\ 
$^9$Be($d,t_0$)$^{8}$Be & $^9$Be($d,\alpha$)$^7$Li  &   &    \\
5.58~\cite{yan,yan_prc} & 20.8~\cite{yan,yan_prc} & 0.26$^{a}$ & 2.3  \\ 
 & & & \\
\multicolumn{2}{c}{$12$}  &  &   \\
$^{10}$B($d,\alpha$)$^8$Be & $^{10}$B($d,p$)$^{11}$B  &   &    \\
102.79~\cite{yan,yan_prc} & 156.1~\cite{yan,yan_prc} & 0.66 & 0.42  \\ 
 & & & \\
\multicolumn{2}{c}{$13$}  &    &    \\
$^{11}$B($d,\alpha$)$^9$Be & $^{11}$B($d,p$)$^{12}$B  &   &    \\
93.2~\cite{yan,yan_prc} & 105.~\cite{yan,yan_prc} & 0.89 & 1.72  \\ 
\end{tabular}
\end{ruledtabular}

\vspace*{.6cm}
\noindent
$^{a}$ Experimental data for the reaction $^9$Be($d,t_0$)$^{8}$Be is measured only for the 
$^8$Be ground state.  
\vspace*{.6cm}
\noindent
\end{table}
\begin{table}
\caption{Ratios of S-factors for the radiative capture reactions to the transfer reactions with identical entrance channels.}
\label{tab:b}
\begin{center}
\begin{ruledtabular}
\begin{tabular}{cccc}
 \multicolumn{2}{c}{$A_{CN}$}      & &   \\
 Reaction1 & Reaction2    &  Ratio & Ratio   \\
 $S_0$(MeVb) & $S_0$(MeVb) & (data) & (calc.) \\
\hline
 \multicolumn{2}{c}{$4$}  &   &    \\
  $d$($d,\gamma$)$t$ & $d$($d,n$)$^3$He  &   &    \\
0.005$\times$10$^{-6}$~\cite{nacre} & 0.0524~\cite{daacv} & 9.5$\times$10$^{-8}$ & 7.9$\times$10$^{-8~a}$  \\ 
 & & & \\
 \multicolumn{2}{c}{$7$}  &   &    \\
 $^6$Li($p,\gamma$)$^7$Be & $^6$Li($p,\alpha$)$^3$He   &   &    \\
 1.07$\times$10$^{-4}$~\cite{nacre} & 3.~\cite{nacre} & 3.6$\times$10$^{-5}$ & 9.6$\times$10$^{-5}$  \\ 
 & & & \\
\multicolumn{2}{c}{$8$}  &   &    \\
$^7$Li($p,\gamma \alpha$)$^4$He & $^7$Li($p,\alpha$)$^4$He   &   &    \\
0.0015~\cite{nacre} & 0.074~\cite{nacre} & 2.0$\times$10$^{-2}$ & 1.4$\times$10$^{-3~b}$   \\ 
 & & & \\
\multicolumn{2}{c}{$10$}  &   &    \\
$^9$Be($p,\gamma$)$^{10}$B & $^9$Be($p,d$)$^8$Be  &   &    \\
0.001~\cite{nacre} & 18.~\cite{nacre} & 5.6$\times$10$^{-5}$ & 6.2$\times$10$^{-5}$ \\ 
$^9$Be($p,\gamma$)$^{10}$B & $^9$Be($p,\alpha$)$^6$Li  &   &    \\
0.001~\cite{nacre} & 18.~\cite{nacre} & 5.6$\times$10$^{-5}$ & 5.9$\times$10$^{-5}$ \\ 
 & & & \\
\multicolumn{2}{c}{$11$}  &   &    \\
$^{10}$B($p,\gamma$)$^{11}$C & $^{10}$B($p,\alpha$)$^{7}$Be  &   &    \\
0.00170~\cite{nacre} & 70.~\cite{nacre} & 2.4$\times$10$^{-5}$ &   5.2$\times$10$^{-5}$ \\ 
 & & & \\
\multicolumn{2}{c}{$12$}  &    &    \\
$^{11}$B($p,\gamma$)$^{12}$C & $^{11}$B($p,\alpha$)$^{8}$Be  &   &    \\
0.0035~\cite{nacre} & 200.0~\cite{nacre} & 1.8 $\times$10$^{-5}$  & 2.4$\times$10$^{-5}$  \\ 
 & & & \\
 \multicolumn{2}{c}{$16$}  &    &    \\
$^{15}$N($p,\gamma$)$^{16}$O & $^{15}$N($p,\alpha$)$^{12}$C  &   &    \\
0.064~\cite{nacre} & 65.~\cite{nacre} & 9.8$\times$10$^{-4}$  & 1.8$\times$10$^{-5}$  \\ 
\end{tabular}
\end{ruledtabular}

\end{center}
\vspace*{.6cm}
\noindent
$^{a}$He($\gamma,d$)D has the photo disintegration cross section 3.2 [$\mu$ b] at the peak of $T_{k\gamma}$=29MeV~\cite{shima:044004} \\
$^{b}$In the reaction $^7$Li($p,\gamma \alpha$)$^4$He we have assumed that $\gamma$ kinetic energy is equivalent to 
the CN temperature.

\vspace*{.6cm}
\noindent
\end{table}
In the $A_{CN}=11$ case in the Table~\ref{tab:a} the model gives a ratio much larger than data for the reactions
$^9$Be($d,t$)$^8$Be/ $^9$Be($d,p$)$^{10}$Be, however we note that the experimental value for the tritium case 
refers to the $t_0$ channel case only.  According to our 
approach, we would estimate an astrophysical factor for the t-channel at least a factor 10 than what is 
quoted in \cite{yan}. For the same
CN case but for the $\gamma$ channel in the Table~\ref{tab:b}, we reproduce the ratio within 
a factor of two.  This is surprising because this reaction is dominated by a 
resonance at about 9 KeV and we have estimated the experimental ratios from the extrapolation of ref.\cite{nacre}.  Evidently the effects of the resonance cancel out by taking ratios. 
The only discordance between a calculated ratio and data is seen in the case of $A_{CN}=8$ in Table~\ref{tab:b}. 
This disagreement of the obtained ratio arises from the fact that
the reaction $^{7}$Li($p,\alpha$)$^{4}$He does not have $l=$0 partial wave in the entrance channel, 
because of the parity conservation in the exit channel. This reaction is restricted to have odd number partial 
waves in the entrance channel.     
As we mentioned in the introductory part, our model developed in this paper assumes $s$-wave
entrance channel for simplicity. Therefore our model fails to estimate the ratio fot this case properly.
We stress that since the entrance channel is the same when we take the ratios, an eventual screening 
should cancel out exactly. Thus our method should give some constraint on extracted bare astrophysical factors.

Another interesting example is D($t,n$)$^4$He/D($^3$He,$~p$)$^{4}$He: the ratios of reactions with similar Q-values.
Using eqs.~(\ref{eq:sfN}-\ref{eq:sfW}) and assuming that the correction coming from the CN decay is given by 
their respective Q-values and temperatures, the ratio of the S-factor is expressed as~\cite{clayton}:
\begin{equation}
  \label{eq:sfRN}
  R=\frac{\mu_i}{\mu_j}\frac{\exp\left({\frac{4}{\hbar} \sqrt {2\mu_j Z_{T_j}Z_{P_j} R_{N_j}}}\right)}{\exp\left({\frac{4}{\hbar}  \sqrt {2\mu_i Z_{T_i}Z_{P_i} R_{N_i}}}\right)}\frac{\Pi_j}{\sum_j \Pi_j}\frac{\sum_i \Pi_i}{ \Pi_i}
\end{equation}
where $i$ and $j$ refer to the different entrance channels.  
For the present case, both reactions have the only main exit channel, so that 
the statistical terms are simplified.
The ratio obtained by Eq.~(\ref{eq:sfRN}), which is shown in the 5th column in the 
Table~\ref{tab:dt}, is smaller than data by a factor of 3. 
If we substitute $\alpha_{n_j}$ by 2.5, we can reproduce the experimental ratio.  
\begin{table}
\caption{Ratios of S-factors for transfer reactions with similar Q-values.}
\label{tab:dt}
\begin{center}
\begin{ruledtabular}
\begin{tabular}{cccc}
 \multicolumn{2}{c}{$A_{CN}$}      & &   \\
 Reaction1 & Reaction2    &  Ratio & Ratio   \\
 $S_0$(MeVb) & $S_0$(MeVb) & (data) & (calc.) \\
\hline
 \multicolumn{2}{c}{$5$}      &  &    \\
 D($t,n$)$^4$He & $^3$He($d,p$)$^{4}$He  &   &    \\
 10.~\cite{nacre} & 6.3~\cite{barker} & 1.6 & 0.42   \\ 
\end{tabular}
\end{ruledtabular}

\end{center}

\end{table}

Last examples we consider the ratios of reactions with different isospins in the entrance channel 
but the same exit channel. 
We consider two ratios for the reactions listed in Table~\ref{tab:difen}. 
All the possible exit channels(i.e., $n,p,d,t,^3$He and $\alpha$) are included to evaluate Eq.~(\ref{eq:sfRN}).  
In the 5th column the ratios obtained by Eq.~(\ref{eq:sfRN}) are shown.
For the first ratio Eq.~(\ref{eq:sfRN}) underestimates the data by a factor of 1000, while for the second ratio
Eq.~(\ref{eq:sfRN}) gives a value in agreement with the data.  
The disagreement of the obtained ratio for the reactions which include lithium isotopes again is caused by the absence 
of the $l=$0 partial wave in the entrance channel of the reaction $^{7}$Li($p,\alpha$)$^{4}$He.  
By contrast, in the case of boron involved reactions both reactions have the $l=$0 component~\cite{barker,ajz} 
and as a matter of fact our model gives reasonable ratio for these two reactions.   
If we use the the $l=1$ term of the formula~(\ref{eq:sfN3}) for the reaction $^{7}$Li($p,\alpha$)$^{4}$He, 
we obtain 9.0, instead of 0.25, as the ratio.    
We can reproduce the ratio by 
substituting $\alpha_{n_j}$ by 3.3 in the former case. It means that the nuclear interaction radius is larger for the 
entrance channel $^6$Li+$d$, which has a dominant $s$-wave component~\cite{barker}, of a factor of 3.3 with respect to the reaction $^7$Li+$p$.

\begin{table}
\caption{Ratios of S-factors for transfer reactions with different entrance channels and identical exit channels.}
\label{tab:difen}
\begin{ruledtabular}
\begin{tabular}{cccc}
 \multicolumn{2}{c}{$A_{CN}$}      & &   \\
 Reaction1 & Reaction2    &  Ratio & Ratio   \\
 $S_0$(MeVb) & $S_0$(MeVb) & (data) & (calc.) \\
\hline
 \multicolumn{2}{c}{$8$}      &     &    \\
 $^{6}$Li($d,\alpha$)$^4$He & $^{7}$Li($p,\alpha$)$^{4}$He  &   &    \\
 16.9~\cite{thm2} & 0.055~\cite{la01} & 307. & 0.25   \\ 
 & & & \\
 \multicolumn{2}{c}{12}      &     &    \\
 $^{10}$B($d,\alpha$)$^{8}$Be  & $^{11}$B($p,\alpha$)$^8$Be  &   &    \\
 102.7~\cite{yan,yan_prc} & 208.0~\cite{aerrss} & 0.49 & 0.46    \\ 
\end{tabular}
\end{ruledtabular}
\end{table}
These two last cases suggest that nuclear effects play an important role and should be considered in detail. 

In conclusion, we have discussed the possible physical origins of the astrophysical S-factors.  We have shown that the
essential contributions to the S-factor 
come from the Coulomb and nuclear penetrability of the colliding nuclei 
to form a CN and from its decay in a given channel. 
We have assumed that this decay process is described statistically.   
By taking suitable ratios of astrophysical factors, one can treat those two ingredients separately 
and compare the model predictions easily to the experimental data.  
In the case where the 'entrance channels' effects are cancelled out in the ratio,
we have shown that our model can reproduce the experimental data without any fitting parameter.
When we consider the cases where the same CN is formed in different reactions, the 
nuclear effects become important. 
When we take the S-factor ratio of a radiative capture reaction to a transfer reaction,
this approach can predict the wide reduction of the yield in the radiative capture. 
Our model demonstrates clearly that 
this reduction arises from the ratio of the $\gamma$ kinetic energy and the ion mass.
Our estimations have been performed in the limit of low incident energies, so that our results refer 
to ratios of bare astrophysical factors at zero incident energy. At such low energies 
experimental data of S-factors are known to be influenced by surrounding electrons.  
 The procedure of taking the ratios of two reactions with an identical entrance channel allows us 
to give some constraint on extracted bare astrophysical factors. 
Having acquired some confidence in our model, we plan
to make use of it to reproduce the experimental data of S-factors as a function of the beam energy.


\bibliography{statS.bib}

\end{document}